\documentclass[12pt,preprint]{aastex}

\newcommand{\be}{\begin{equation}}
\newcommand{\ee}{\end{equation}}
\newcommand{\vrel}{{ v_{\rm rel}}} 
 
\newcommand{\rhogas}{{ \rho_{\rm gas}}}
\newcommand{\cs}{{ c_{\rm s}}}
\newcommand{\bmin}{{ b_{\rm min} }}
\newcommand{\kon}{{ K }} 

\begin{document} 

\title{\bf MIGRATION OF EXTRASOLAR PLANETS: \\ EFFECTS FROM X-WIND ACCRETION DISKS}

\author{Fred C. Adams$^{1,2}$, Mike J. Cai$^{1,3}$, and Susana Lizano$^{1,4}$} 

\affil{$^1$Michigan Center for Theoretical Physics \\
Physics Department, University of Michigan, Ann Arbor, MI 48109} 

\affil{$^2$Astronomy Department, University of Michigan, Ann Arbor, MI 48109} 

\affil{$^3$Academia Sinica, Institute of Astronomy and Astrophysics, Taiwan}

\affil{$^4$Centro de Radioastronom{\'i}a y Astrof{\'i}sica, UNAM, 58089 
Morelia, Michoac{\'a}n, Mexico} 

\begin{abstract} 

Magnetic fields are dragged in from the interstellar medium during the
gravitational collapse that forms star/disk systems. Consideration of
mean field magnetohydrodynamics (MHD) in these disks shows that
magnetic effects produce subkeplerian rotation curves and truncate the
inner disk. This letter explores the ramifications of these predicted
disk properties for the migration of extrasolar planets. Subkeplerian
flow in gaseous disks drives a new migration mechanism for embedded
planets and modifies the gap opening processes for larger planets. This
subkeplerian migration mechanism dominates over Type I migration for
sufficiently small planets ($m_P\lesssim{1}M_\earth$) and/or close
orbits ($r\lesssim{1}\rm{AU}$).  Although the inclusion of
subkeplerian torques shortens the total migration time by only a
moderate amount, the mass accreted by migrating planetary cores is
significantly reduced.  Truncation of the inner disk edge (for typical
system parameters) naturally explains final planetary orbits with
periods $P\sim4$ days. Planets with shorter periods $P\sim2$ days can
be explained by migration during FU-Ori outbursts, when the mass
accretion rate is high and the disk edge moves inward. Finally, the
midplane density is greatly increased at the inner truncation point of
the disk (the X-point); this enhancement, in conjunction with
continuing flow of gas and solids through the region, supports the in
situ formation of giant planets.

\end {abstract} 


\newpage 

\section{INTRODUCTION}

The population of observed extrasolar planets exceeds 350, with
many objects found in short-period orbits \cite{udry}. Tremendous
progress has been made in our theoretical understanding of the
migration mechanisms that allow planets to move (usually inward) as
they form (Papaloizou \& Terquem 2006, hereafter PT06). Parallel
advances have been made in our understanding of star formation
processes \cite{bo}, which produce the circumstellar disks that
provide birth places for planets. However, these latter developments
have not been fully integrated into theories of planet
formation/migration. This letter explores the implications of the
predicted disk properties for planet migration.

During the collapse phase of star formation, magnetic fields are
dragged inward and retained by the nascent disk.  These magnetic
fields, along with those generated by the star, can drive powerful
outflows (Blandford \& Payne 1982, Shu et al. 1994), provide channels
for accretion of disk material onto the star \cite{bouvier}, and
produce a back reaction on the structure (Shu et al. 2007, hereafter
S07). In particular, circumstellar disks are predicted to display
subkeplerian rotation curves.  We parameterize the rotation speed of
the gas through the ansatz $\Omega_{\rm gas}=f\Omega$, where
$\Omega^2=GM_\ast/r^3$.  Recent MHD calculations indicate that
$f\approx0.66$ for disks around T Tauri stars and $f\approx0.39$ for
disks undergoing FU-Orionis outbursts (S07).  The relative velocity
between the gas and orbiting planets is thus a substantial fraction of
the Keplerian rotation speed, and $f\ne1$ over much of the disk.

Planetary migration is altered when the circumstellar gas moves at
subkeplerian speeds.  In Keplerian disks, embedded planets drive wakes
into the disk, and asymmetric gravitational interactions of these
wakes drive planets inward \cite{gt79}, a process called Type~I
migration (Ward 1997ab). In subkeplerian disks, embedded planets orbit
with Keplerian speeds and experience a headwind from the gaseous disk
(which moves slower).  This velocity mismatch results in energy loss
from the orbit and inward migration. This paper explores this
mechanism (Section 2), including an assessment of the conditions
necessary for it to compete with Type~I migration.
A weaker version of this effect occurs in nearly Keplerian disks,
where $1-f={\cal O}(H^2/r^2)\sim1/400$, where $H$ is the disk scale
height. The headwind produced by this small departure from Keplerian
flow affects dust grains as they coagulate into larger bodies, but
does not affect planets \cite{stu2}. The departures from Keplerian
speeds due to magnetic effects are larger by two orders of magnitude.

The second feature of magnetic disks is that their inner edge is
truncated. Disk material accretes onto the star by moving along
magnetic field lines in a ``funnel flow''.  In one theory, the disk
truncation point, the start of the funnel flow, and the origin point
for outflows occur at a location known as the X-point (Cai et al.
2008, hereafter C08).  This ``point'' has a finite extent, comparable
to the disk scale height. In competing theories, the wind originates
from an extended disk region \cite{pudnorm}, but the disk retains an
inner truncation point; disk wind models also predict subkeplerian
flow, which might inhibit wind launching (Shu et al. 2008). Because
the disk does not extend to the star, migration is naturally halted at
the truncation point $(r\sim{0.05}\rm{AU})$, as recognized
\cite{lin96} soon after the discovery of extrasolar planets (Mayor \&
Queloz 1995, Marcy \& Butler 1996).  This paper argues that observed
ultra-short-period planets ($P\sim{1-2}\rm{day}$, $r\sim{0.02}\rm{AU}$) 
can be understood through disk truncation at the X-point combined with
FU-Ori outbursts (Section 3).

\section{SUBKEPLERIAN MIGRATION} 

This section presents a migration mechanism driven by subkeplerian
flow in magnetically controlled disks.  To illustrate the properties
of this mechanism, we consider power-law disks, where the surface
density and temperature distributions take the form
$\Sigma(r)=\Sigma_1(r_1/r)^p$ and $T(r)=T_1(r_1/r)^q$.  We take
$r_1=1\rm{AU}$, so the coefficients $\Sigma_1$ and $T_1$ correspond to
values at $1\rm{AU}$. The power-law index for the surface density lies 
in the range $1/2<p<2$ \cite{cassen}. For the Minimum Mass Solar Nebula
(MMSN), the surface density constant $\Sigma_1\approx4500$~g/cm$^2$
\cite{kuchner}. The power-law index for the temperature profile lies
in the range $1/2<q<3/4$. Viscous accretion disks \cite{pringle} and
flat reprocessing disks \cite{as86} have $q\approx3/4$; flared
reprocessing disks \cite{chiang} and the early solar nebula
\cite{stu1} have $q\approx1/2$.
Planetary orbits are taken to be circular, with orbital angular
momentum $J=m_P(GM_\ast r)^{1/2}$. The disk scale height
$H=\cs/\Omega$, where $\cs$ is the sound speed, which is determined by
the disk temperature profile. The radial dependence of $H/r$ is thus
$H/r=(H/r)_1(r/r_1)^{(1-q)/2}$, where $(H/r)_1\approx1/20$ (S07).

\subsection{Torques} 

For circular orbits, the torque $T_X$ exerted on the planet by gas in
the subkeplerian disk takes the form
\be{T_X=C_D{\pi}R_P^2r\rhogas\vrel^2,}\label{torque}\ee 
where $r$ is the orbital radius, $\vrel$ is the relative velocity
between the gas and planet, $\rhogas$ is the gas density, and $R_P$ is
the planetary radius.  The dimensionless drag coefficient $C_D\sim1$
for planets \cite{stu2}, but can be enhanced by gravitational focusing
(Ohtsuki et al. 1988).  The gas density $\rhogas\approx\Sigma/2H$.
Because the gaseous disk orbits with subkeplerian speeds 
$\Omega_{\rm gas}=f\Omega$, the relative velocity takes the form
$\vrel=(1-f)(GM_\ast/r)^{1/2}$, and the torque becomes
\be{T_X}={\pi\over2}C_D(1-f)^2GM_\ast{R_P}^2\left({\Sigma\over{H}}\right).\ee

The time evolution of the semi-major axis obeys the differential equation 
\be{1\over{r}}{dr\over{dt}}={2T_X\over{m_P}\Omega{r^2}},\ee
where we assume that the orbit remains circular. The 
migration timescale is given by 
\be{t_0}\equiv{1\over{\pi}C_D(1-f)^2}\left({H\over{r}}\right)\left({m_P\over\Sigma R_P^2}\right){1\over\Omega}.\ee
For typical parameters, $M_\ast=1.0M_\odot$, $m_P=1.0M_\earth$, 
$R_P=1.0R_\earth$, $\Sigma_1=4500$~g/cm$^2$, $f=0.66$, and $(H/r)_1=0.05$, 
the timescale $t_0\approx70,000\rm{yr}$. 

\begin{figure} 
\figurenum{1}
{\centerline{\epsscale{0.90} \plotone{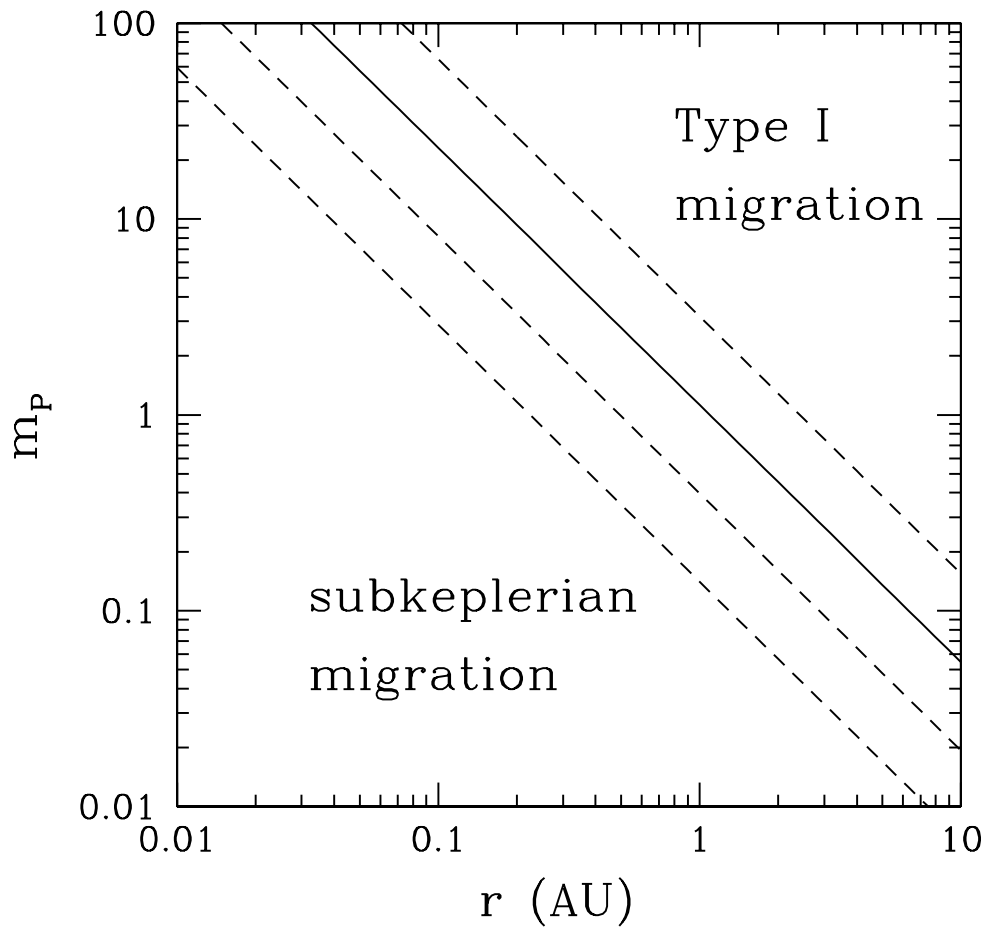} } } 
\figcaption{Regions of $(r,m_P)$ parameter space for which migration 
from subkeplerian flow (with $f=0.66$) dominates Type~I migration.
Planetary mass ($M_\earth$) is plotted versus orbital radius (AU). 
Subkeplerian migration dominates in the region below the lines, 
which are drawn for stellar masses $M_\ast/M_\odot=2,1,0.5,0.25$ 
(top to bottom). }
\label{fig:mrplane} 
\end{figure}

For comparison, Type~I torque strengths are given by 
\be{T_I}=C_I\left({m_P\over{M_\ast}}\right)^2{\pi}\Sigma r^2(r\Omega)^2\left({r\over{H}}\right)^2,\ee
where the dimensionless constant ${C_I}\approx(1.364+0.541p)/\pi\approx0.606$ 
(Ward 1997a, Tanaka 2002). The ratio of the torque $T_X$ to the Type~I 
torque is given by
\be{T_X\over{T_I}}=\left[{C_D(1-f)^2\over{2}{C_I}}\right]\,
\left({M_\ast\over{m_P}}\right)^2\,\left({R_P\over{r}}\right)^2\,
\left({H\over{r}}\right)\,\propto\,M_\ast^2\,m_P^{-4/3}\,r^{-(3+q)/2}.\label{ratio}\ee
For an Earth-like planet ($m_P=1M_\earth$, $R_P=1R_\earth$) orbiting a
solar mass star, $C_D=C_I$, and $H/r=1/20$, the two torques have
almost the same magnitude at $r=1\rm{AU}$ when $(1-f)=0.34$.  
Since the torque ratio grows with decreasing semi-major axis,
subkeplerian torques $T_X$ are stronger for $r\lesssim1\rm{AU}$. The
ratio decreases with the planetary mass, so that larger planets are
more affected by Type~I migration. This torque ratio also depends on
the stellar mass, so that subkeplerian torques are more important for
systems associated with more massive stars. Figure \ref{fig:mrplane}
shows the regions of parameter space $(r,m_P,M_\ast)$ for which
subkeplerian migration dominates over Type I migration.  Subkeplerian
torques are relatively stronger in the inner disk, where magnetic
effects ($f\ne1$) are most likely to arise.

Note that this treatment is approximate: The $T_I$ expression neglects
magnetic effects, which can be important in these systems (Terquem
2003). In general, $T_I$ should be a function of $f$.  Moreover, this
Type I torque is derived for isothermal disks, whereas X-wind disks
have temperature gradients, which can affect migration (Paardekooper
\& Mellema 2006).  Finally, the two torques $(T_I,T_X)$ are not fully
independent --- both mechanisms produce disk wakes that can interfere
with each other. The comparison shown in Figure \ref{fig:mrplane} is
thus subject to future corrections.

\subsection{Gaps} 

In standard treatments of gap formation, parcels of gas flowing past
the planet are scattered by its gravitational potential, and deflected
through the angle $\theta=(2Gm_P/b\vrel^2)$, where $b$ is the impact
parameter of the encounter (PT06). This deflection changes the
parcel's specific angular momentum by the increment
\be\Delta{J}={2\kon{G^2}m_P^2r\over{b^2}\vrel^3},\ee
where the constant $\kon$ includes corrections arising from the
rotating reference frame (Goldreich \& Tremaine 1980).  The disk
angular momentum changes at a rate determined by the integral of
$\Delta{J}$.

In the Keplerian case, the planet moves faster (slower) than disk
material on its outside (inside). The planet thus transfers angular
momentum to the outer disk and negative angular momentum to the inner
disk, and opens a gap from both directions. However, in a subkeplerian
disk, the planet moves faster than the disk material on both sides of
the potential gap; this complication modifies the gap formation
process.

Although supersonic flow ($\vrel\ge\cs$) alters this picture, we
start by estimating the angular momentum transferred by ``scattering''
gas parcels and compare the result to that transferred by viscosity.
Consider the disk exterior to the planetary orbit. The planet moves
faster than the gas, and angular momentum is transferred outward at
the rate
\be{dH\over{dt}}=\int_{\bmin}^\infty(\Delta J)\Sigma\vrel{db}=
2\kon G^2m_P^2r\int_{\bmin}^\infty{\Sigma db\over{b^2}\vrel^2}
\approx{2\kon\over(1-f)^2}{Gm_P^2r^2\Sigma\over\bmin M_\ast},
\label{hdotscatter}\ee
where the integral's support is concentrated (by assumption) near the
minimum impact parameter $\bmin$. This expression neglects disk shear
terms, which must be included to recover the standard result in the
limit $f\to1$ (compare equation [\ref{hdotscatter}] with PT06).

Gap formation requires that the rate of angular momentum transfer to
the disk from ``scattering'' exceeds the rate at which angular
momentum is transferred by viscosity. This condition requires
\be{dH\over{dt}}\gtrsim{3}\pi\nu\Sigma{r}^2\Omega,\label{hdotviscous}\ee 
where the viscosity $\nu$ is parameterized as $\nu=(2/3)\alpha\cs{H}$,
and the viscosity coefficient $\alpha$ is dimensionless.

Gap formation requires a second condition: The planet's sphere 
of influence must be large enough, e.g., the Roche radius 
$R_R=r(m_P/3M_\ast)^{1/3}$ must exceed the scale height $H$. The
requirement $R_R>H$ places another constraint on the planet mass
required for gap opening:
\be{m_P\over{M_\ast}}\ge{3}\left({H\over{r}}\right)^3.\label{gaproche}\ee
This condition allows the flow to experience nonlinear deviations
\cite{ward1}, including shock formation in the vicinity of the planet
\cite{kory}. Since this requirement could be modified for high Mach
numbers, equation (\ref{gaproche}) should be considered as approximate.
This constraint allows us to use $\bmin=R_R\propto{r}$ to evaluate the
angular momentum transfer rate in equation (\ref{hdotscatter}), and
solve equation (\ref{hdotviscous}) to find the planet mass required
for gap opening:
\be\left({m_P\over{M_\ast}}\right)^{5/3}\ge{3^{2/3}\pi(1-f)^2\over{2}\kon}{\nu\over\Omega r^2} 
\approx\pi(1-f)^2\,\alpha\left({H\over{r}}\right)^2.\label{gapscatter}\ee
Gaps form when both constraints (\ref{gaproche}) and
(\ref{gapscatter}) are satisfied (see also Crida et al. 2006).  
For $\alpha\approx{10^{-3}}$, $f=0.66$, and $H/r={1/20}$, 
the nonlinear constraint (\ref{gaproche}) and the viscous constraint
(\ref{gapscatter}) are comparable, and the mass required for gap
clearing is $\sim{100}M_\earth$ (about one Saturn mass).  Because the
planet moves faster than the gas on both sides of the gap, the planet
readily transfers angular momentum to the outer gap edge, but tends
to draw disk material outward from the inner edge. When the planet has
sufficient mass (equation [\ref{gaproche}]), this material shocks
against the wake of the planet. The planet will thus be located closer
to the inner gap edge.

\subsection{Orbital Evolution} 

To consider orbital evolution of the planet, we define a dimensionless
angular momentum variable $\xi\equiv{J}/J_1$, where $J=m_P\Omega{r}^2$
and $J_1$ is the orbital angular momentum at $r=1\rm{AU}$.
We also define reduced torque parameters through the ansatz
$\Gamma_I\equiv{T_I{(J_1)}}/J_1$ and $\Gamma_X\equiv{T_X{(J_1)}}/J_1$.
The equation of motion then becomes 
\be{d\xi\over{dt}}=-\left[{\Gamma_I}\xi^{-a}+{\Gamma_X}\xi^{-b}\right],\ee
where the indices $a=2(p-q)$ and $b=2p+3-q$. For typical values
$p=3/4$ and $q=1/2$, the torque indices become $a=1/2$ and $b=4$.
Since the two torques are nearly equal for Earth-mass planets at $1\rm{AU}$, 
$\Gamma_X\approx\Gamma_I\approx{10}({\rm Myr})^{-1}=1/(0.1\rm{Myr})$.  
Keep in mind that both torque expressions are approximate (see Section 
2.1). 

For subkeplerian torques acting alone, the time required to migrate from a
starting location $\xi_0=J_0/J_1=(r_0/1\rm{AU})^{1/2}$ to the central
star is given by ${\Gamma_X}{t_X}=\xi_0^{b+1}/(b+1)$. Similarly, the
time required for Type~I torques to move the planet from $\xi_0$ to
the star is given by ${\Gamma_I}{t_I}=\xi_0^{a+1}/(a+1)$. Since we
expect $b\gg{a}$, subkeplerian migration is much faster than Type~I
migration for planets starting inside $\sim1\rm{AU}$. For planets
starting at larger radii, however, Type~I torques dominate until the
planet reaches $r\sim1\rm{AU}$, where subkeplerian torques take over.

The above results assume that the planetary core mass remains
constant. However, the core mass must increase with time for giant
planets to form, and the torque strengths change as the mass
increases. To illustrate this behavior, we use a simplified treatment
of planetesimal accretion where the planetary core grows according to
\be{dm_P\over{dt}}=C_{\rm in}\pi{R_P}R_R\Sigma_Z\Omega,\ee
where $\Sigma_Z\propto\Sigma$ is the surface density in solids
(Papaloizou \& Terquem 1999, Laughlin et al. 2004a). Giant planet
formation typically requires 
$\Sigma_Z\approx130\,\,{\rm g}\,{\rm cm}^{-2}$ at $r=1\rm{AU}$,
somewhat larger than the MMSN value (Lissauer \& Stevenson 2007).  
For a given density of planetary cores, 
$\rho_C\approx{3.2}\,\,{\rm g}\,{\rm cm}^{-3}$, 
the planetary radius $R_P$ is determined by the mass.  
After defining a scaled radius $\eta\equiv{R_P}/R_\earth$, 
we write the core accretion process in dimensionless form,
\be{d\eta\over{dt}}={\gamma}\xi^{-(2p+1)},\label{radgrow}\ee
where $\gamma\approx{20}(\rm{Myr})^{-1}$. We can combine planetary
growth with orbital evolution. As the planet grows according to
equation (\ref{radgrow}), the torques vary with planetary radius, 
and the equation of motion for $\xi\equiv(r/{1}\rm{AU})^{1/2}$ becomes
\be{d\xi\over{dt}}=-\left[{\hat\Gamma}_I\eta^3\xi^{-a}+
{\hat\Gamma}_X\eta^{-1}\xi^{-b}+3\gamma\eta^{-1}\xi^{-2p}\right],\label{eqmotion}\ee
where $\hat{\Gamma}_I=\Gamma_I(\eta=1)$ and 
$\hat{\Gamma}_X=\Gamma_X(\eta=1)$.

\begin{figure} 
\figurenum{2}
{\centerline{\epsscale{0.90} \plotone{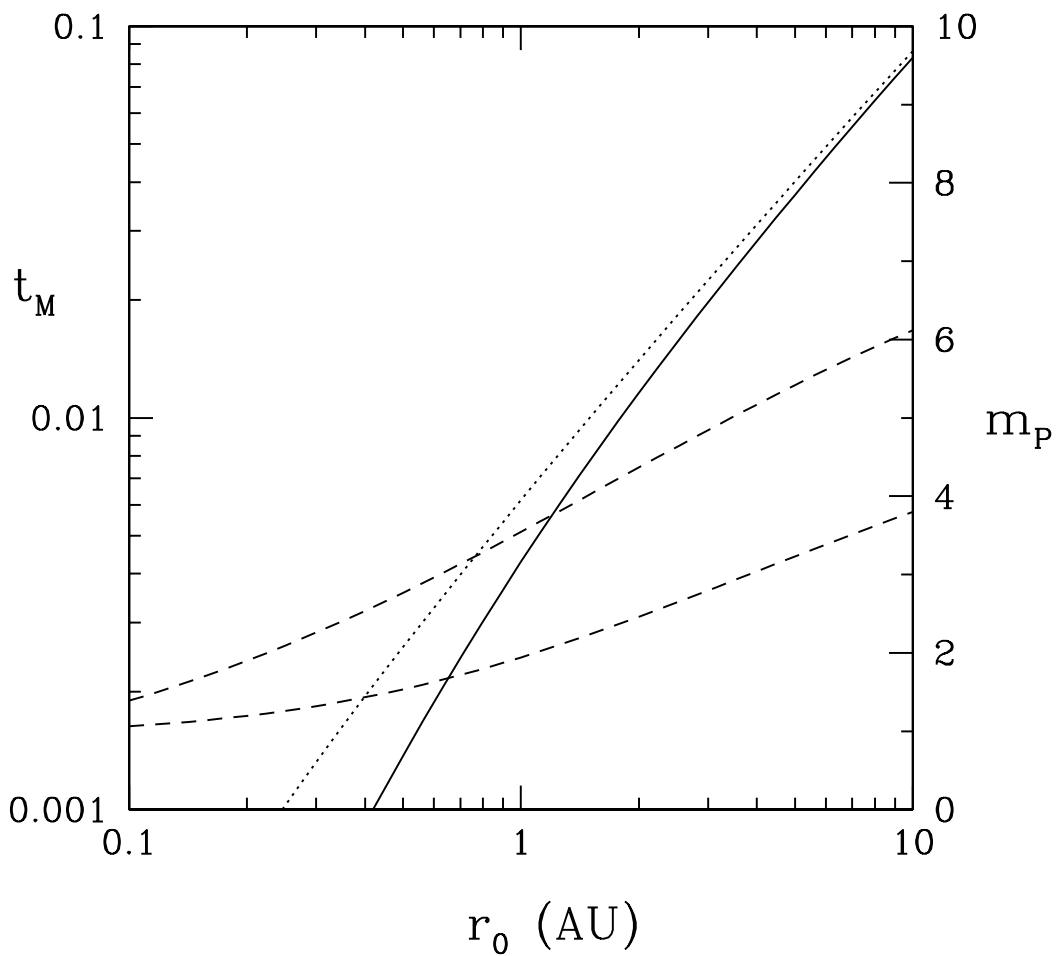} } } 
\figcaption{Migration time and final core mass versus starting radius.
Solid curve shows migration time $t_M$ (Myr, left axis) including
subkeplerian and Type I torques. Dotted curve shows $t_M$ for Type I
torques only. Dashed curves show final core mass $m_P$ ($M_\earth$,
right axis) for migration with both torques (bottom) and Type I
torques only (top). }
\label{fig:orbit} 
\end{figure}

When migration reaches $r\lesssim{1}\rm{AU}$, subkeplerian torques
become larger than Type~I torques and planetary migration takes place
more rapidly.  If migration starts at larger radii, however, most of
the evolution time is spent at larger radii where Type~I migration
torques prevail. Consequently, subkeplerian torques don't
substantially change the total migration time for starting radii
beyond $r\sim{2}\rm{AU}$.  However, the mass accreted during the
migration epoch is significantly altered due to acceleration of the
latter stages. These trends are illustrated by Figure \ref{fig:orbit},
which shows migration times and final masses for planets starting at
varying locations $r_0$ with fixed initial mass $m_P={1}M_\earth$.
Here, the equations of motion (\ref{radgrow},\ref{eqmotion}) are
integrated using the standard parameter values quoted above.

For completeness, we note that subkeplerian flow occurs only in
magnetically active disk regions. The flow becomes Keplerian in 
``dead zones'' \cite{gammie}, where this migration mechanism will not
operate. However, ordinary Type~I migration continues to act within
dead zones, so that migration occurs at the standard Type~I rate.

\section{DISK TRUNCATION} 

This section considers the implications of disk truncation.  In X-wind
theory, this inner cutoff radius is given by
\be{r_X}=\left[{\mu_\ast^4\over{G}M_\ast{\dot M}_D^2\Phi_{\rm dx}^4}\right]^{1/7},\label{xpoint}\ee
where $\mu_\ast$ is the magnetic dipole moment of the star, $M_\ast$
is the stellar mass, ${\dot M}_D$ is the mass accretion rate, and
$\Phi_{\rm dx}\approx{1}$ is a dimensionless parameter (Ostriker \&
Shu 1995, S07).  Because $r_X$ marks the inner disk boundary, planets
cannot migrate inside the X-point through standard migration
mechanisms involving disk torques because there is no disk material.
The X-point location ${r_X}\approx{0.05}\rm{AU}$ for typical
parameters (C08), and is coincident with the semi-major axes of 
``hot Jupiters''.  Disk truncation through magnetic effects thus
provides a natural mechanism for ending migration and explains the
observed planets in $\sim4$-day orbits. However, extrasolar planets
have recently been discovered with significantly smaller semi-major
axes\footnote{Schneider, J. 2009, Extrasolar Planets Encyclopedia,
http://exoplanet.eu/catalog-all.php}, $r\approx{0.02}\rm{AU}$. 
Although some variation in $r_X$ is expected, the presence of
extrasolar planets in these extremely tight orbits represents 
some tension with expectations from X-wind theory, and suggests 
the possibility of an additional mechanism. 

The presence of these short-period planets can be understood as
follows: The location of the X-point depends on the mass accretion
rate through the disk. During FU-Ori outburst events, the mass
accretion rate is greatly enhanced \cite{hartmann}, and the X-point
moves inward (S07). Equation (\ref{xpoint}) indicates that $\dot{M}_D$
must increase by a factor $\sim{25}$ to move the X-point to
${r_X}\approx{0.02}\rm{AU}$ as required. Since outburst phases have
$\dot{M}_D\approx{10^{-4}}{M_\odot}{\rm yr}^{-1}$, planets can (in
principle) migrate inward during these FU-Ori phases.

This simple picture has two complications. First, FU-Ori outbursts are
relatively short-lived. The time-averaged mass accretion rate is
$\dot{M}_D\sim10^{-7}{M_\odot}\rm{yr}^{-1}$.  During outbursts, the
$\dot{M}_D$ is $\sim{100}$ times larger, so that the outburst duty
cycle is correspondingly shorter, and each event lasts only $100-1000$
years.  During the outburst phase, the disk extends inward to the
desired radii, and migration can take place.  However, the planets
will not necessarily have enough time to migrate inward far enough
during the (short) outburst phases. This issue is made more urgent by
the timing of the events: The planets take a relatively long time to
form and migrate inward to the ``standard'' location of the X-point
$(3-10\rm{Myr})$.  Thus, relatively few FU-Ori outbursts will take
place after the planets arrive at $r=r_X\approx{0.05}\rm{AU}$.

On the other hand, migration will be aided by the fact that the
planets might not have time to open up gaps in the disk. As a result,
we expect that the migration rates will be faster than the standard
Type~II migration rates (for giant planets with gaps). The migration
rate will be closer to the Type~I migration rates, which are fast, and
are even faster for larger mass planets (that usually open up a gap).

The second complication is that, because FU-Ori outbursts are short,
the star has difficulty adjusting its rotation period to co-rotate with
the disk: During the quiescent phase, the star rotates every $\sim4$
days, in sync with the inner disk edge at $r\approx{0.05}\rm{AU}$.
During an outburst, the large accretion rate suddenly pushes the disk
edge to $r\approx{0.02}\rm{AU}$.  For the star to co-rotate with the
new disk edge, it needs four times its angular momentum. However, the
maximum rate for the star to gain angular momentum is 
$\dot{J}\approx\dot{M}_D{r_X}^2{\Omega}_X$. For
$\dot{M}_D={10^{-4}}M_\odot\rm{yr}^{-1}$ and $r_X=0.02\rm{AU}$, 
it takes $2000-3000$ years to spin up the star, somewhat longer than
the timescale for each outburst.  As a result, the inner disk edge
generally rotates faster than the star, and the magnetic funnel forms
a leading spiral.  Angular momentum flows into the star and makes the
truncation radius smaller than its equilibrium value (equation
[\ref{xpoint}]). These departures from equilibrium thus allow
migrating planets to move even farther inward.

Next we note that the density of gaseous material in the disk midplane
is enhanced over the expected value $\rho\approx\Sigma/2H$. Because
disk truncation by the funnel flow increases the density by a factor
${r}/H\sim{40}$ at the X-point \cite{shu94}, the expected density
$\rho_X\approx{r}\Sigma/2H^2$.  This density enhancement has important
ramifications:

[1] Giant planets that successfully migrate to the X-point at
$r\approx0.05\rm{AU}$ can be pushed further inward (perhaps to
$r\approx0.02\rm{AU}$) during FU-Ori outburst events. With the density
enhancement at the disk midplane, the timescale for inward movement
can be short ($\lesssim{40}\rm{yr}$), so that migration can take place
during a single outburst.

[2] Planetary cores can migrate inward to the X-point on a short
timescale, often fast enough that they cannot become gas giants before
they reach the inner disk edge. If planetary cores are large enough,
they continue to accrete gaseous material while they orbit at the
X-point. The density enhancement considered herein provides a
corresponding enhancement in the supply of available material. One
issue associated with forming hot Jupiters in situ is the lack of
gaseous material in the inner disk. This gas shortage is largely
alleviated for planets at the X-point, both from this density
enhancement and from the continuing flow of material into the region
(Ward 1997b).  On the negative side, high temperatures at the X-point
make the gas essentially dust-free. Gas accretion onto planets is
limited by cooling processes, which will be compromised if the opacity
is too low.

\section{CONCLUSION} 

This letter presents a new migration mechanism for embedded planets.
These bodies experience a headwind due to the subkeplerian rotation
curve of the gas in magnetically controlled disks, and drag forces
drive inward migration. This mechanism dominates over Type I migration
(Figure \ref{fig:mrplane}) for sufficiently small planets
($m_P\lesssim{1}M_\earth$) and/or close orbits ($r\lesssim{1}\rm{AU}$).  
The total migration time moderately decreases due to subkeplerian
torques, but the mass accreted by planetary cores during the migration
epoch changes more substantially (Figure \ref{fig:orbit}).

The gap opening process is modified in subkeplerian disks because the
planet moves faster than the gas (supersonically) both inside and
outside its orbit. Gap opening thus requires nonlinear interactions
(approximately given by equation [\ref{gaproche}]). The estimated mass
threshold for gap opening is ${m_P}\ge{100}M_\earth$, and the planet
resides near the inner gap edge.

For typical parameters, magnetic truncation of the inner disk occurs
at $r_X\approx0.05\rm{AU}$, and thus naturally explains observed
extrasolar planets in $\sim4$-day orbits.  Migration of planets to
smaller orbits, as sometimes observed, requires an additional
mechanism. During FU-Ori outbursts, the X-point moves inward to the
required location $r_X\approx0.02\rm{AU}$ due to the large accretion
rate, which also allows fast inward migration. X-wind theory 
(S07,C08) thus provides a viable explanation for all of the currently
observed short-period planets (long-period planets require additional 
considerations). 

Finally, planetary cores that halt migration at the X-point
($r_X\approx0.05\rm{AU}$) can be subjected to a significantly enhanced
gas density and a continuing supply of material (including inwardly
migrating solids).  These features help proto-hot-Jupiters finish
forming in their observed $\sim4$-day orbits.

This letter shows that magnetic effects drive subkeplerian migration
and produce final orbits with $r\approx{r_X}\approx0.02-0.05\rm{AU}$
(from disk truncation). These processes must compete with other
mechanisms acting on embedded planets, including Type I migration
(Ward 1997ab), stochastic migration from turbulence (Nelson \&
Papaloizou 2004, Laughlin et al. 2004b, Nelson 2005, Johnson et
al. 2006, Adams \& Bloch 2009), runaway migration (Masset \&
Papaloizou 2003), torques exerted by toroidal fields (Terquem 2003),
opacity transitions (Menou \& Goodman 2004), and radiative
(non-barotropic) effects (Paardekooper \& Mellema 2006, Paardekooper
\& Papaloizou 2008).  Future work should study how these processes
interact with each other to ultimately form giant planets.

\acknowledgments

We thank Frank Shu and Greg Laughlin for discussions, and MCTP for
hospitality while this work was developed. FCA is supported by NASA
Grant:NNX07AP17G and NSF Grant:DMS-0806756; MJC by
Grant:NSC-95-2112-M-001-044; and SL by Grant:CONACyT-48901.

\end{document}